\RequirePackage{lineno}
\documentclass[prd,twocolumn,amsmath,amssymb]{revtex4-1}

\usepackage{overpic}
\usepackage{graphicx}
\usepackage{epsfig}
\usepackage{dcolumn}
\usepackage{bm,color}
\usepackage{rotating}
\usepackage{multirow}
\usepackage{booktabs}
\usepackage{appendix}
\usepackage{subfigure}
\usepackage{hhline}
\usepackage{lineno}
\usepackage{mathrsfs}
\usepackage{verbatim}
\usepackage{amsmath}
\usepackage{amssymb}
\usepackage{amsfonts}
\usepackage{amsbsy}
\usepackage{float}
\usepackage{enumerate}
\usepackage{enumitem}

\usepackage{setspace} 
\usepackage{hyperref}
\hypersetup{colorlinks,linkcolor=blue,anchorcolor=blue,citecolor=blue,urlcolor=blue}

\let\oldequation\equation
\let\oldendequation\endequation

\renewenvironment{equation}
  {\linenomathNonumbers\oldequation}
  {\oldendequation\endlinenomath}

\newcommand{\mev}{\mathrm{MeV}}

\newcommand{\gev}{\mathrm{GeV}}
\newcommand{\gevc}{\mathrm{GeV}/c}
\newcommand{\gevcc}{\mathrm{GeV}/c^2}
\newcommand{\psip}{\psi(2S)}
\newcommand{\etap}{\eta'}
\newcommand{\EE}{e^+e^-}
\newcommand{\LL}{l^+l^-}
\newcommand{\jpsi}{J/\psi}
\newcommand{\pp}{\pi^+\pi^-}
\newcommand{\ppz}{\pi^0\pi^0}

\def \ifb  {\mbox{fb$^{-1}$}}

\begin{document}

\title{\boldmath Search for $e^{+}e^{-}\to\etap\psi(2S)$ at center-of-mass energies from 4.66 to 4.95 GeV}

\author{
\begin{small}
\begin{center}
M.~Ablikim$^{1}$, M.~N.~Achasov$^{4,c}$, P.~Adlarson$^{76}$, O.~Afedulidis$^{3}$, X.~C.~Ai$^{81}$, R.~Aliberti$^{35}$, A.~Amoroso$^{75A,75C}$, Q.~An$^{72,58,a}$, Y.~Bai$^{57}$, O.~Bakina$^{36}$
% ...
\\
\vspace{0.2cm}
(BESIII Collaboration)\\
\vspace{0.2cm} {\it
$^{1}$ Institute of High Energy Physics, Beijing 100049, People's Republic of China\\
% ...
$^{3}$ Bochum  Ruhr-University, D-44780 Bochum, Germany\\
$^{4}$ Budker Institute of Nuclear Physics SB RAS (BINP), Novosibirsk 630090, Russia\\
% ...
$^{35}$ Johannes Gutenberg University of Mainz, Johann-Joachim-Becher-Weg 45, D-55099 Mainz, Germany\\
$^{36}$ Joint Institute for Nuclear Research, 141980 Dubna, Moscow region, Russia\\
% ...
$^{57}$ Southeast University, Nanjing 211100, People's Republic of China\\
$^{58}$ State Key Laboratory of Particle Detection and Electronics, Beijing 100049, Hefei 230026, People's Republic of China\\
% ...
$^{72}$ University of Science and Technology of China, Hefei 230026, People's Republic of China\\
% ...
$^{75}$ University of Turin and INFN, (A)University of Turin, I-10125, Turin, Italy; (B)University of Eastern Piedmont, I-15121, Alessandria, Italy; (C)INFN, I-10125, Turin, Italy\\
$^{76}$ Uppsala University, Box 516, SE-75120 Uppsala, Sweden\\
% ...
$^{81}$ Zhengzhou University, Zhengzhou 450001, People's Republic of China\\
\vspace{0.2cm}
$^{a}$ Deceased\\
% ...
$^{c}$ Also at the Novosibirsk State University, Novosibirsk, 630090, Russia\\
% ...
}
\end{center}
\vspace{0.4cm}
\end{small}
}


\begin{abstract}
Using data samples with an integrated luminosity of 4.67~$\mathrm{fb}^{-1}$ collected by the BESIII detector operating at the BEPCII collider, we search for the process $e^+e^- \rightarrow \eta' \psi(2S)$ at center-of-mass energies from $4.66$ to $4.95~\mathrm{GeV}$. No significant signal is observed, and upper limits for the Born cross sections $\sigma^B(e^+e^-\rightarrow\eta'\psi(2S))$ at the 90\% confidence level are determined.
\end{abstract}
\maketitle
\oddsidemargin -0.2cm
\evensidemargin -0.2cm


\section{\boldmath Introduction}

The charmonium sector with bound states of charm and anti-charm quarks, provides an ideal laboratory for exploring the interplay between perturbative and nonperturbative effects in Quantum Chromodynamics (QCD)~\cite{Brambilla:2014jmp}. Since the discovery of the $\chi_{c1}(3872)$ in 2003~\cite{Belle:2003nnu}, numerous charmonium-like states have been observed. After their properties do not align with those expected for conventional charmonium states, they are promising candidates for exotic hadrons.
The vector states (labelled with $\psi(\mathrm{mass})$)~\cite{BaBar:2005hhc, BaBar:2006ait, BESIII:2016adj} are directly produced in $\EE$ annihilations or initial state radiation (ISR) processes.

The $\psi(4660)$ was initially reported by the Belle experiment and later by BaBar in the ISR process $e^+e^- \to \gamma_{\rm ISR} \pi^+ \pi^- \psi(2S)$~\cite{Belle:2007umv,BaBar:2012hpr}. BESIII measured the $\EE \to \pi^+\pi^-\psip$ cross sections at center-of-mass (c.~m.) energies ($\sqrt{s}$) between $4.0076$ and $4.6984~\gev$ using 35 scan samples with a total integrated luminosity of $20.1~\ifb$~\cite{BESIII:2021njb}. The results are consistent with those of the earlier BaBar and Belle experiments, but with significantly improved precision. The BESIII collaboration has measured a mass of $(4651.0\pm 37.8 \pm 2.1)~\mev$ and width of $(155.4\pm24.8\pm0.8)~\mev$ for the $\psi(4660)$. 
Using the ISR method, the Belle experiment identified a structure close to the $\psi(4660)$ in the $\EE \to \Lambda_c\bar{\Lambda}_c$ cross-section shape~\cite{Belle:2008xmh}. Measured by the BESIII collaboration, the cross section of the same process reveals a different line shape~\cite{BESIII:2017kqg}. Various theoretical interpretations have been proposed to explain the nature of the $\psi(4660)$, such as hybrid meson~\cite{Chen:2016qju}, molecular state~\cite{Guo:2010tk}, hadrocharmonium~\cite{Dubynskiy:2008mq}, and tetraquark state~\cite{Cotugno:2009ys}. 
Besides the $\psi(4660)$, two structures around $4.75~\gev$ were reported in recent studies of $\EE\to K\bar{K}\jpsi$~\cite{BESIII:2023wqy,BESIII:2022kcv} and $\EE\to D_{s}^{*+}D_{s}^{*-}$~\cite{BESIII:2023wsc} processes at BESIII. Further exploration of the decay properties is essential for a comprehensive understanding of the internal structure of these vector structures.

In heavy $Q\bar{Q}$ systems, the hadronic transitions serve as a crucial probe of their internal structures and help to establish the understanding of light quark coupling with a heavy degree of freedom~\cite{Anwar:2016mxo}. 
The hadronic transition between two vector quarkonia via an $\eta$ meson is suppressed by $SU(3)$ flavor symmetry and heavy quark spin symmetry~\cite{Voloshin:2011hw} compared to the transition producing two pions. 
A notable example for this is the low transition ratio between $\Gamma(\psi(2S)\to J/\psi\eta)$ and $\Gamma(\psi(2S)\to J/\psi\pi\pi)$. Inspired by the Nambu-Jona-Lasinio model, the authors of Ref.~\cite{Anwar:2016mxo} estimated the partial widths $\Gamma(Y\to J/\psi\eta)$ of $\psi(4360)$, $\psi(4390)$, and $\psi(4660)$ by assuming them as $c\bar{c}$ bound states. 
The $\EE\to\eta\jpsi$~\cite{BESIII:2023tll}, $\EE\to\etap\jpsi$~\cite{BESIII:2019nmu}, 
and $\EE\to\eta\psi(2S)$~\cite{BESIII:2024jzt} processes have been
studied by the BESIII experiment. Here, the $\psi(4040)$, $\psi(4230)$, and $\psi(4360)$ were observed in the $\eta\jpsi$ process. An enhancement around $4.2~\gev$ is found in the cross section of $\EE\to\etap\jpsi$. Using data samples collected between $\sqrt{s}=4.288~\gev$ and $4.951~\gev$, the $\EE\to\eta\psip$ process is observed with a statistical significance above $5.0\sigma$. The $\EE\to\etap\psip$ process has not been investigated yet.

In this article, we study the $e^+e^- \to \etap \psi(2S)$ process and investigate possible vector charmonium or charmonium-like states in its cross section.
The used data samples were collected at nine different c.~m.~energy points ranging from $4.66$ to $4.95~\gev$ between 2020 and 2021, with a total integrated luminosity of $4.67$~\ifb. 
The $\psip$ is reconstructed via the pion transition process $\psi(2S) \to \pi^+ \pi^- J/\psi$, and $J/\psi \to e^+e^-$ or $ \mu^+\mu^-$. The $\etap$ is reconstructed by two decay modes $\etap\to\gamma\pp$ (Mode I) and $\etap\to\eta\pp$ with $\eta\to\gamma\gamma$ (Mode II) to collect more statistics.

\section{\boldmath Experiment setup and MC simulation}
The BESIII detector~\cite{Ablikim:2009aa} records symmetric $e^{+}e^{-}$ collisions provided by the BEPCII storage ring~\cite{Yu:IPAC2016-TUYA01} in the c.~m. energy range $ 2.0 < \sqrt{s} < 4.95~\gev$, with a peak luminosity of $1 \times 10^{33}\;\text{cm}^{-2}\text{s}^{-1}$ achieved at $\sqrt{s} = 3.77\;\text{GeV}$. BESIII has collected large data samples in this energy region~\cite{Ablikim:2019hff,EventFilter}. The cylindrical core of the BESIII detector covers 93\% of the full solid angle and consists of a helium-based multilayer drift chamber~(MDC), a plastic scintillator time-of-flight system~(TOF), and a CsI(Tl) electromagnetic calorimeter~(EMC), which are all enclosed in a superconducting solenoidal magnet providing a 1.0~T magnetic field. The solenoid is supported by an octagonal flux-return yoke with resistive plate counter muon identification modules interleaved with steel. The charged-particle momentum resolution at $1~{\rm GeV}/c$ is $0.5\%$, and the ${\rm d}E/{\rm d}x$ resolution is $6\%$ for electrons from Bhabha scattering. The EMC measures photon energies with a resolution of $2.5\%$ ($5\%$) at $1$~GeV in the barrel (end cap) region. The time resolution in the TOF barrel region is 68~ps, while that in the end cap region was 110~ps. The end cap TOF system was upgraded in 2015 using multigap resistive plate chamber technology, providing a time resolution of 60~ps.

Monte Carlo (MC) simulated events are used to determine the detection efficiency, optimize selection criteria, and investigate possible background contributions. These samples are produced with a {\sc geant4}-based~\cite{geant4} package, which includes the geometric description of the BESIII detector and the detector response.
The simulation models the beam energy spread and initial state radiation in the $e^+e^-$ annihilations with the generator {\sc kkmc}~\cite{ref:kkmc}.
The inclusive MC sample includes the production of open charm processes, the ISR production of vector charmonium(-like) states, and the continuum processes incorporated in {\sc kkmc}~\cite{ref:kkmc}. 
All particle decays are modeled with {\sc evtgen}~\cite{ref:evtgen} using branching fractions either taken from the Particle Data Group (PDG)~\cite{ParticleDataGroup:2022pth}, when available, or otherwise estimated with {\sc lundcharm}~\cite{ref:lundcharm}. Final state radiation~(FSR) from charged final state particles is incorporated using the {\sc photos} package~\cite{photos}.
The signal MC samples of $e^{+}e^{-} \to \etap \psip $ are generated for each c.~m.~energy assuming the Born cross sections following the line shape of the $\psi(4660)$, which is parameterized with a Breit-Wigner function using the mass and width taken from the PDG~\cite{ParticleDataGroup:2022pth}.

\section{\boldmath Signal selection and background modeling}
The signal process of $e^{+}e^{-} \to \etap \psip$ comprises six charged tracks and one photon for Mode I or two photons for Mode II. 
Charged tracks detected in the MDC must be within a polar angle ($\theta$) range of $|\rm{cos\theta}|<0.93$, where $\theta$ is the polar angle with respect to the positive direction of the $z$-axis, the symmetry axis of the MDC.
For charged tracks, the distance of closest approach to the interaction point (IP) must be less than 10\,cm along the $z$-axis, and less than 1\,cm in the transverse plane.
Photon candidates are identified by EMC showers. The deposited energy of each shower must be more than 25~MeV in the barrel region where $|\rm cos \theta|< 0.80$ and more than 50~MeV in the end cap regions where $|\rm cos \theta|$ is between $0.86$ and $0.92$. To exclude showers originating from the charged tracks, the angle subtended by the EMC shower and the position of the closest charged track at the EMC must be greater than 10 degrees as measured from the IP. To suppress electronic noise and showers unrelated to the event, the difference between the EMC time and the event start time is required to be within [0, 700]\,ns.

To enhance the event selection efficiency, we include events with five charged tracks in our analysis, which constitute about $32\%$ and $37\%$ according to MC simulation for Mode I and Mode II, respectively.
A candidate event should meet either of the following criteria: it must consist of five charged tracks with a net charge of $\pm 1$, accounting for the possibility of one missing charged pion. Alternatively, it should involve six charged tracks with a net charge of zero. In Mode I, there must be at least one photon, and Mode II requires at least two photons.

MC simulations indicate a distinct kinematic separation between pions and leptons for signal events.
The charged tracks with momentum above $1.0~\gevc$ are categorized as leptons while those with momenta below $0.8~\gevc$ are classified as pions. Electrons and muons are distinguished using the ratio of their energies~($E$) deposited in the EMC to momenta~($p$): the charged tracks are labeled as electrons if $E/p > 0.7$, and as muons if $E/p < 0.7$ and $E < 0.45~\gev$. 

In order to further suppress background contributions and improve the mass resolution, a four-constraint~(4C) kinematic fit is performed for six-track events by requiring the four-momenta of the final state particles to be consistent with the initial one. For five-track events that account for a missing pion, a kinematic fit with one degree of freedom (1C) is applied by constraining the mass of the missing particle to the pion nominal mass.
The optimal photon(s) is(are) those that yield the minimum $\chi^2$ value. The four-momenta of the final state particles used in the following analysis are from the kinematic fit.
The corresponding $\chi^2$ values are required to satisfy $\chi^2_{4\rm C} < 30$~($\chi^2_{4\rm C} < 60$) or $\chi^2_{1\rm C} < 10$~($\chi^2_{1\rm C} < 25$) for Mode I~(Mode II), respectively.  These criteria are optimized by maximizing ${\epsilon}/(a/2+\sqrt{B})$, where $\epsilon$ denotes the signal MC efficiency, $B$ is normalized number of background events based on the inclusive MC sample, and $a$ is the number of $\sigma$ corresponding to one-sided Gaussian tests at the desired significance level~\cite{Punzi:2003bu}.
In Mode I,
to suppress the background contribution from decays with unexpected number of photons, such as  $e^+e^- \to \pi^+\pi^-J/\psi$ with $J/\psi$ decaying to the same final states.We perform a 4C kinematic fit, requiring $\chi^2_{4\rm{C},n\gamma} < \chi^2_{4\rm{C},(n-1)\gamma}$. Here $\chi^2_{4\rm{C},n\gamma}$ is obtained from a 4C kinematic fit including n photons expected for the signal candidate, while $\chi^2_{4\rm C,(n-1)\gamma}$ is determined from an additional 4C fit with one fewer photon compared to the signal decay.
This requirement results in an efficiency loss of $9.6\%$, while $69.7\%$ of background events can be removed. This requirement is not applied in Mode II due to the low background level.

The signal candidates are required to be within the $J/\psi$ mass region, defined as $[3.083,3.111]~\gevcc$ and $[3.073,3.121]~\gevcc$ for the invariant mass of lepton pairs in Mode I and Mode II, respectively. The mass requirement of the $\eta$ candidate is set to be $[0.482,0.604]~\gevcc$ in Mode II.
The mass requirements of the $\jpsi$ and $\eta$ candidates are optimized by maximizing $\epsilon/(a/2+\sqrt{B})$.
This analysis involves four pion tracks in the final states, yielding four potential combinations of $\psi(2S)$ candidates for each event. A mass window of $[3.680,3.693]~\gevcc$ around the $\psip$ peak is applied to the combinations of invariant mass $M(\pp J/\psi)$. In order to improve the mass resolution, the invariant mass is calculated using $M(\pp\LL) - M(\LL) + m(J/\psi)$, where $m(J/\psi)$ is the nominal mass of $J/\psi$.
For those candidates that survived, the one with minimal value of
$|\frac{M(\gamma\pi^+\pi^-/\eta\pp) - m(\etap)}{\sigma_{\etap}}|^{2} + |\frac{ {|M(\pi^+\pi^-J/\psi) - m(\psi(2S))}}{\sigma_{\psi(2S)}}|^{2}$ 
is selected. Here $m(\etap)$ and $m(\psi(2S))$ represent the nominal masses of $\etap$ and $\psi(2S)$ quoted from the PDG~\cite{ParticleDataGroup:2022pth}, while $\sigma_{\etap}$ and $\sigma_{\psi(2S)}$ are the mass resolutions obtained from signal MC simulation of the corresponding invariant mass distributions.

After applying the above selection criteria, the possible remaining background contributions are studied based on the inclusive MC samples with an integrated luminosity equivalent to 10 times of the real data.
For Mode I, the dominant remaining background is the $\EE\to\pi^+\pi^-\psip$ and $\EE\to\pi^0\pi^0\psip$ processes. To better estimate their contributions,
100,000 exclusive MC events are generated at each c.~m.~energy point with modeling of the intermediate states and the cross section line shape taken from Ref.~\cite{BESIII:2021njb}. 
For Mode II, the number of the background contribution estimated by using inclusive MC samples 
are $0.6$, $0.8$, $0.6$, $0$, $0.1$, $0.1$, $0$, $0$, and $0$ for the nine data samples, which is negligible.

\section{\boldmath Measurement of the cross section}

The Born cross section for studied signal process is determined by
\begin{equation}\label{formula:cs}
\small
    \sigma^{ B} = \frac{N_{\rm sig}}{\mathcal{L}_{\rm int} \cdot (1+\delta) \cdot \frac{1}{\left| 1-\Pi \right| ^2} \cdot \epsilon \cdot \mathcal{B} },
\end{equation}
where $N_{\rm sig}$ represents the observed signal yield, $\mathcal{L}_{\rm int}$ is the integrated luminosity of the data, $1+\delta$ denotes the radiative correction factor obtained from the {\sc kkmc} generator, $\frac{1}{\left| 1-\Pi \right| ^2}$ stands for the vacuum polarization factor derived from QED calculations~\cite{WorkingGrouponRadiativeCorrections:2010bjp}, $\epsilon$ is the detection efficiency estimated through MC simulation, and $\mathcal{B}$ represents the product of the branching fractions of intermediate states in the decay cascade, quoted from the PDG~\cite{ParticleDataGroup:2022pth}, which is 1.22\% for Mode I and 0.692\% for Mode II.

The Born cross sections are obtained by performing simultaneous unbinned maximum likelihood fits to the corresponding $M(\gamma\pp)$ and $M(\eta \pp)$ mass spectra, where the Born cross section value for $e^+e^- \to \etap \psi(2S)$ is shared between the two decay modes at the same c.~m.~energy.
The signal shapes for both modes are modeled by MC-simulated shapes.
In Mode I, the background shapes originating from $e^{+} e^{-} \rightarrow \pi\pi\psi(2S)$ processes are taken from the invariant mass spectrum of the corresponding simulated MC samples. Other background contributions are described by the invariant mass spectrum of the inclusive MC samples after excluding the aforementioned processes. 
Regarding Mode II, only the signal component is considered since the background contribution is negligible.
This assumption results in a conservative estimation of the upper limit of the Born cross section, as well as the upper limit of $\Gamma_{ee}\cdot\mathcal{B}$ which will be introduced later.
In the fitting procedure, the number of $e^{+} e^{-} \rightarrow \pi\pi\psi(2S)$ events is fixed to the values determined by using the corresponding cross section and 
detection efficiency, and the number of other background contributions 
is a free parameter.
The fit results for the data sample collected at $4.78~\gev$ are shown in Fig.~\ref{fig:fit1}, 
the results for other energy points can be found in Appendix~\ref{appendix}.
Due to low statistics, the upper limits on the cross section at the 90\% confidence level (C.~L.) are provided using the profile likelihood method incorporating the systematic uncertainties~\cite{Stenson:2006gwf}. 
The cross sections and signal yields are listed in Table~\ref{tab:result} for the different c.~m.~energies. 
Since we do not allow negative signal yields due to low statistics, the error bars on data points corresponding to zero events do not represent two-sided standard 68\% confidence level.

\begin{table*}[htbp]
\centering
\caption{The numerical results for $\EE\to\etap\psip$. $\mathcal{L}$ is the integrated luminosity, $(1+\delta)$ is the ISR correction factor, $\frac{1}{|1-\Pi|^2}$ is the VP correction factor, $\epsilon_{1}$ and $\epsilon_{2}$ are the selection efficiencies for Mode I and Mode II respectively, $N_{\rm sig}$ denotes the number of signal events determined from the simultaneous fit, $\sigma^B$ represents Born cross sections, $\alpha$ is the significance of the signal events, $N^{\rm ul}$ denotes the upper limits of signal events, and $\sigma^{\rm ul}$ is the upper limit of the Born cross section.}
\label{tab:result}
\begin{tabular}{ccccccccccc}
\hline\hline
$\sqrt{s} (\mev)$& $\mathcal{L}({\rm pb}^{-1})$& $(1+\delta)$& $\frac{1}{|1-\Pi|^2}$& $\epsilon_{1}(\%)$& $\epsilon_{2}(\%)$& $N_{\rm sig}$& $\sigma^B$(pb)& $\alpha$& $N^{\rm ul}$& $\sigma^{\rm ul}$(pb)\\
\hline
4661.24& 529.43& 0.764& 1.054& 25.6& 21.6& $0.0^{+0.7}_{-0.0}$&    $0.0^{+0.5}_{-0.0}$&  0.0& 3.3& 2.5\\
4681.92& 1667.39& 0.865& 1.054& 22.8& 20.5& $0.0^{+1.9}_{-0.0}$&    $0.0^{+0.4}_{-0.0}$&   0.0& 6.4& 1.5\\
4698.82& 535.54& 0.935& 1.055& 21.4& 19.3& $2.9^{+1.8}_{-1.4}$&    $2.1^{+1.3}_{-1.0}$&   1.3& 7.1& 5.1\\
4739.70& 163.87& 1.092& 1.055& 18.6& 17.0& $1.1^{+1.2}_{-0.7}$&    $2.6^{+2.8}_{-1.6}$&   1.4& 4.0& 9.3\\
4750.05& 366.55& 1.133& 1.055& 18.0& 16.3& $0.0^{+0.5}_{-0.0}$&    $0.0^{+0.5}_{-0.0}$&   0.0& 2.5& 2.6\\
4780.54& 511.47& 1.241& 1.055& 16.4& 14.4& $2.8^{+1.8}_{-1.3}$&    $2.1^{+1.3}_{-1.0}$&  0.1& 7.7& 5.7\\
4843.07& 525.16& 1.438& 1.056& 14.0& 11.8& $1.6^{+1.4}_{-0.9}$&    $1.2^{+1.0}_{-0.7}$&  1.7& 5.0& 3.7\\
4918.02& 207.82& 1.664& 1.056& 11.5& 9.8& $0.9^{+1.2}_{-0.7}$&    $1.8^{+2.3}_{-1.4}$&  0.8& 3.7& 7.2\\
4950.93& 159.28& 1.771& 1.056& 10.8& 8.7& $0.0^{+0.7}_{-0.0}$&    $0.0^{+1.8}_{-0.0}$&  0.0& 4.3& 10.9\\
\hline\hline
\end{tabular}
\end{table*}

\begin{figure}[htbp]
    \centering
        \includegraphics[width=0.5\textwidth]{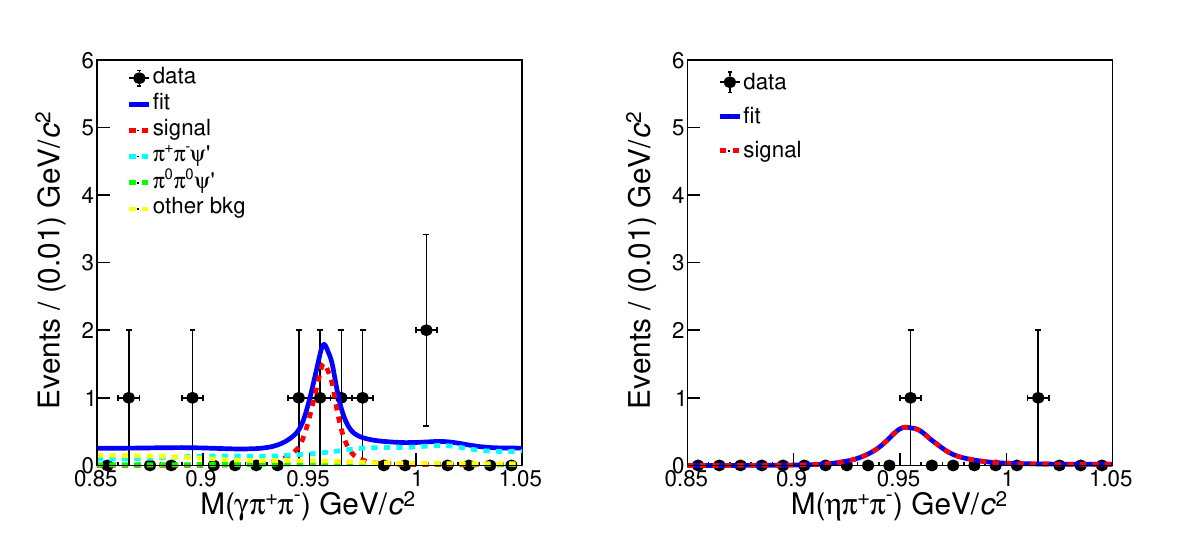}
    \caption{
    Distributions of $M(\gamma\pp)$ for Mode I (left) and $M(\eta\pp)$ for Mode II (right) for data at a c.~m.~energy of $4.78~\gev$. 
    The black dots with error bars are the data, the solid blue lines denote the best fit results, the red dashed lines represent the possible signal contribution, and the dashed curves in other colors (cyan, green, and yellow) indicate the background contributions.
    }
    \label{fig:fit1}
\end{figure}

To study possible charmonium-like states in the $\EE\to\etap\psip$ cross section, a maximum likelihood fit is performed to the dressed cross section. The likelihood is defined as
\begin{equation}
\small
L\left(\mu^{\mathrm{sig}} ; p\right)=\prod_{i=1}^{9} L_i\left(\mu_{i}^{\mathrm{sig}} ; p_i\right),
\end{equation}
where $\mu_{i}^{\text {sig }}$ is the expected number of signal events and $p_i$ are the parameters of the fit models. $L_i$, as a function of $\mu_{i}^{\text {sig }}$, is obtained by scanning the number of $\etap$ signals in the simultaneous fit at each c.~m.~energy point.
The systematic uncertainties of the Born cross section, which will be discussed in Section~\ref{sec:v}, have been considered in the construction of $L_i$. 
Assuming that the $\etap\psip$ signal originates from one resonance, the dressed cross section is parameterized as a relativistic Breit-Wigner function:
\begin{equation}
\small
        \sigma(m)=\bigg|\frac{\sqrt{12 \pi \mathcal{B} \Gamma_{ee} \Gamma}}{s-M^2+i M \Gamma} \sqrt{\frac{\Phi(m)}{\Phi\left(M\right)}}\bigg|^{2},
\end{equation}
where $\Phi(m)$ is the phase space factor, $M$ and $\Gamma$ represent the mass and width of the resonance, respectively, $\Gamma_{ee}$ denotes the electronic partial width, and $\mathcal{B}$ is the branching fraction to $\etap\psi(2S)$.
Two alternative resonance assumptions of the $\psi(4660)$ and $\psi(4710)$ are used to describe the line shape of the dressed cross section. The mass and width of the $\psi(4660)$ are fixed to the PDG~\cite{ParticleDataGroup:2022pth} values and the ones of $\psi(4710)$ are fixed to the BESIII results~\cite{BESIII:2023wqy}. The fit results are shown in Fig.~\ref{fig:ul3}. The upper limits of $\Gamma_{ee}\cdot  \mathcal{B}$ are determined at the 90\% C.~L.~with the likelihood profile method, and are less than $0.22$~eV and $0.54$~eV for the $\psi(4660)$ and $\psi(4710)$, respectively. 

\begin{figure}[htbp]
\includegraphics[width=0.45\textwidth]{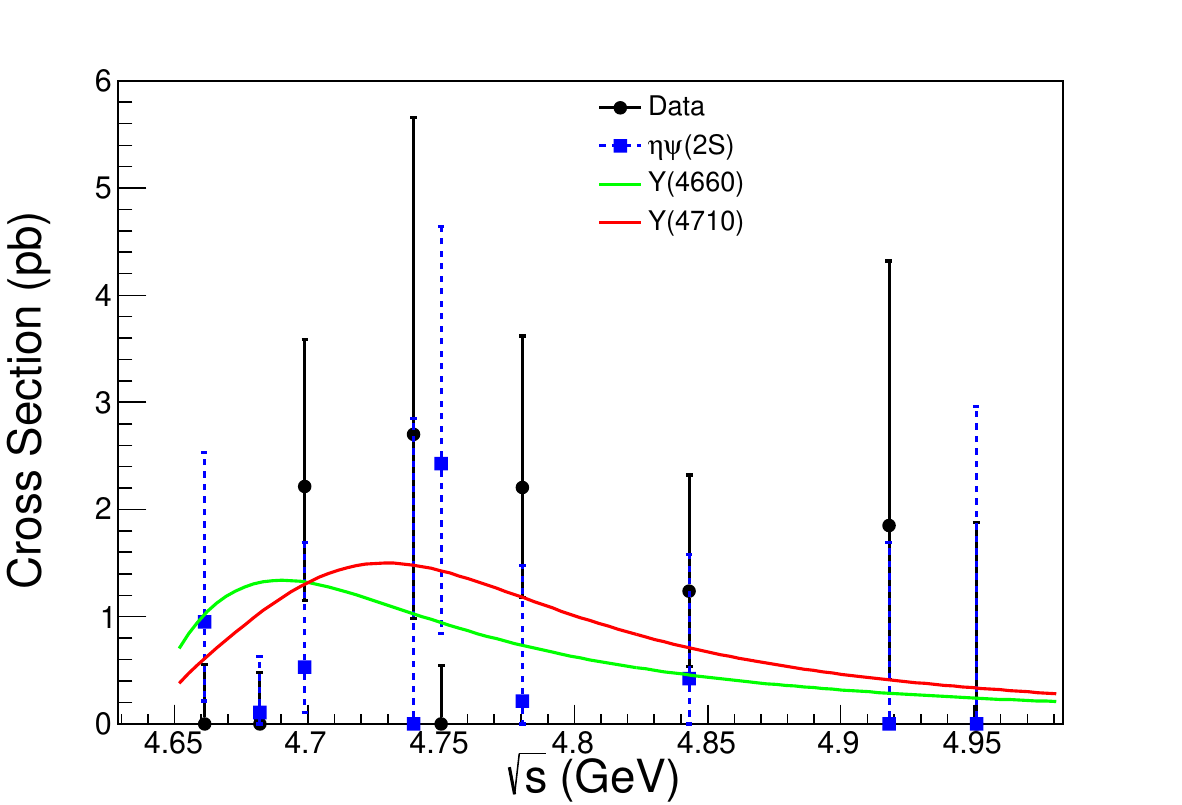}
\caption{Fit to the measured dressed cross section with the Breit-Wigner function. The dots with error bars are the measured cross sections at nine c.~m.~energy points. The green and red lines represent the fit results with the $\psi(4660)$ and $\psi(4710)$ assumptions, respectively. }
\label{fig:ul3}
\end{figure}

\section{\boldmath Systematic uncertainties}~\label{sec:v}

The sources of the systematic uncertainties in the measurements of the Born cross section are classified into multiplicative and additive terms. 

The multiplicative terms include the luminosity, the branching fractions from intermediate states, the radiative correction, the vacuum polarization,  and the detection efficiency. They are described in detail below.

\begin{enumerate}[label=(\arabic*),left=1em] 
  \item The uncertainty of the integrated luminosity is 0.6\% as obtained by analyzing the large-angle Bhabha scattering events~\cite{BESIII:2022ulv}. 
  \item  The input branching fractions of $\psip$ and $\etap$ are quoted from the PDG~\cite{ParticleDataGroup:2022pth}.
  \item  To evaluate the uncertainty of the ISR factor, two sources are considered. 
  First, the input line shape is changed to either a constant term or $q^3/s$, instead of the Breit-Wigner function of the $\psi(4660)$. Here $q^3$ is the P-wave phase space factor with $q$ denoting the momentum of $\eta'$ in the $e^+e^-$ c.~m.~frame.
  Second, considering the uncertainties of the parameters of the $\psi(4660)$, we change the mass and width by one standard deviation.
  The largest difference on the resulting cross section is assigned as the uncertainty.
  \item The uncertainty of the vacuum polarization factor is taken as 0.5\% from the QED calculation~\cite{Burkhardt:1989ky}. 
  \item For photon and pion detection efficiencies, the uncertainties are $1.0\%$ per photon~\cite{BESIII:2010ank} and $1.0\%$ per pion~\cite{Yuan:2015wga}, respectively.  
   The uncertainties of photons are $1.0\%$ for Mode I and $2.0\%$ for Mode II.
   Events with three or four pions are kept and they are combined in each mode according to the detection efficiencies.
   Then the uncertainties from the two modes are combined considering the associated  branching fractions.
    \item The uncertainty of the lepton tracking is estimated to be $1.0\%$ per track~\cite{BESIII:2013ris}. The total uncertainty is $2.0\%$. 
   \item The uncertainties of the 4C and 1C kinematic fit is estimated by correcting the helix parameters of the charged tracks according to the method in Ref.~\cite{BESIII:2012mpj}. The difference in the selection efficiency with or without the correction is taken as the systematic uncertainty.
   \item To evaluate the uncertainty of the mass windows, the MC shape is smeared with a Gaussian function to get better consistency with data. The parameters of the Gaussian function are taken from Ref.~\cite{BESIII:2021njb} for $J/\psi$ and $\psip$, and Ref.~\cite{BESIII:2022yoo} for $\eta$. The difference on the detection efficiencies before and after smearing the Gaussian function is taken as the uncertainty. 
\end{enumerate}

The additive terms originate from the fit to the invariant mass distributions of the $\etap$ candidates, consisting of the signal shape, the background shape, and the background size. 
To estimate the uncertainty from the signal shape arising from the mass resolution difference between data and MC simulation, the MC shape is smeared with a Gaussian function taken from Ref.~\cite{BESIII:2019nmu}. The uncertainty of the background shape is estimated by replacing the inclusive MC shape with a polynomial. The systematic uncertainty from the fixed background contribution from $\EE\to\pp\psip$ and $\EE\to\ppz\psip$ is estimated by varying the expected number of background events by one standard deviation. 
The maximum number of signal events among the above conditions is used to calculate the upper limit of the cross section. Additionally, the multiplicative terms are convoluted with the likelihood distributions in the form of a Gaussian function. 

The systematic uncertainties in two modes are combined according to the branching fractions and detection efficiencies. 
The combined systematic uncertainties are summarized in Table~\ref{tab:sys}. Assuming all sources are independent, the total uncertainties are obtained by adding the individual values in quadrature.

\begin{table*}[hbtp]
\centering
\caption{Relative systematic uncertainties (in unit of \%) at different c.~m.~energies.}
\label{tab:sys}
\setlength{\tabcolsep}{1mm}{

\begin{tabular}{cccccccccc}
\hline\hline
Energy point(MeV)	& 4660& 4680& 4700& 4740& 4750& 4780& 4840& 4914& 4946\\ 
\hline  
Luminosity & 0.6& 0.6& 0.6& 0.6& 0.6& 0.6& 0.6& 0.6& 0.6\\
Branching ratio $\psip$  & 1.0& 1.0& 1.0& 1.0& 1.0& 1.0& 1.0& 1.0& 1.0\\ 
Branching ratio $\etap$ & 1.3& 1.3&  1.3&  1.3&  1.3&  1.3&  1.3&  1.3&  1.3\\
ISR factor 1 &   15.2& 18.1& 17.9& 18.0& 18.1& 18.0& 15.0& 13.0& 13.1\\
ISR factor 2 &  6.7& 5.4& 6.1& 8.1& 8.1& 8.7& 8.3& 8.7& 9.4\\
VP factor & 0.5& 0.5& 0.5& 0.5& 0.5& 0.5& 0.5& 0.5& 0.5\\
Lepton efficiency & 2.0& 2.0& 2.0& 2.0& 2.0& 2.0& 2.0& 2.0& 2.0\\
Pions efficiency &  3.6& 3.6& 3.6& 3.6& 3.6& 3.6& 3.6& 3.6& 3.6\\
Photon efficiency &  1.4& 1.4& 1.4& 1.4& 1.4& 1.4& 1.4& 1.4& 1.4\\
Kinematic fit &   3.0& 2.8& 2.9& 2.7& 2.5& 2.4& 2.3& 2.8& 2.7\\
Mass window $\eta$&  0.3& 0.3& 0.3& 0.3& 0.3& 0.3& 0.3& 0.3& 0.3\\ 
Mass window $J/\psi$ &  0.3& 0.3& 0.3& 0.3& 0.3& 0.3& 0.3& 0.3& 0.3\\
Mass window $\psi(2S)$ & 0.5& 0.5& 0.5& 0.5& 0.5& 0.5& 0.5& 0.5& 0.5\\
\hline
Total& 17.6&19.7&19.7&20.5&20.5&20.7&17.9&16.6&17.0\\
\hline\hline
\end{tabular}
}
\end{table*}

\section{\boldmath Summary}
In summary, by analyzing data samples with an integrated luminosity of $4.67~\ifb$ collected by the BESIII detector operating at the BEPCII collider, the $\EE\to\etap\psip$ process is investigated at c.~m.~energies between $4.66$ and $4.946~\gev$. 
No significant signal is observed from the data, and the upper limits of the Born cross section $\sigma^B(\EE\to\etap\psip)$ at the 90\% C.~L.~ are determined at the nine c.~m.~energy points.
The vector resonance contribution is investigated by the analysis of the dressed cross section. No statistically significant contribution from the vector states  $\psi(4660)$ or $\psi(4710)$ is found. The corresponding upper limit of the product of electronic partial width and branching fraction to $\etap\psip$ at the 90\% C.~L.~are estimated to be smaller than 0.22~eV or 0.54~eV, respectively.
The cross sections of $\EE\to\etap\psi(2S)$ are found to be 
approximately comparable
those of $\EE\to\eta\psi(2S)$~\cite{BESIII:2024jzt}, while the cross section ratio of $\EE\to\etap\jpsi$ to $\EE\to\eta\jpsi$ is about $0.1$ in the $\psi(4230)$ mass region. Future investigations using data samples with higher luminosity, such as the upgraded BEPCII~\cite{BESIII:2020nme,Geng:2021qqj} and STCF~\cite{Achasov:2023gey}, are necessary to provide further insight.

\section{\textbf{Acknowledgement}}

The BESIII Collaboration thanks the staff of BEPCII and the IHEP computing center for their strong support. This work is supported in part by National Key R\&D Program of China under Contracts Nos. 2020YFA0406300, 2020YFA0406400, 2023YFA1606000; National Natural Science Foundation of China (NSFC) under Contracts Nos. 12375070, 11635010, 11735014, 11935015, 11935016, 11935018, 12025502, 12035009, 12035013, 12061131003, 12192260, 12192261, 12192262, 12192263, 12192264, 12192265, 12221005, 12225509, 12235017, 12361141819; the Chinese Academy of Sciences (CAS) Large-Scale Scientific Facility Program; the CAS Center for Excellence in Particle Physics (CCEPP); Joint Large-Scale Scientific Facility Funds of the NSFC and CAS under Contract No. U2032108, No. U1832207; 100 Talents Program of CAS; The Institute of Nuclear and Particle Physics (INPAC) and Shanghai Key Laboratory for Particle Physics and Cosmology; German Research Foundation DFG under Contracts Nos. 455635585, FOR5327, GRK 2149; Istituto Nazionale di Fisica Nucleare, Italy; Ministry of Development of Turkey under Contract No. DPT2006K-120470; National Research Foundation of Korea under Contract No. NRF-2022R1A2C1092335; National Science and Technology fund of Mongolia; National Science Research and Innovation Fund (NSRF) via the Program Management Unit for Human Resources \& Institutional Development, Research and Innovation of Thailand under Contract No. B16F640076; Polish National Science Centre under Contract No. 2019/35/O/ST2/02907; The Swedish Research Council; U. S. Department of Energy under Contract No. DE-FG02-05ER41374

\appendix
\section{\boldmath The $\etap$ mass distribution at other c.~m.~energies}
\label{appendix}



The fitting outcomes for each energy point and the combined data are presented in Fig.~\ref{fig:fit2}. The combined data fitting, incorporating the background component in Mode II and modeled as a 1st-order polynomial, yields a value of $N_{\rm sig} = 7.2^{+3.9}_{-3.4}$ with a statistical significance of $2.1$ and an upper limit at $N^{\rm ul}=12.8$. 

\begin{figure*}[htbp]
    \centering
        \subfigure[~$4.66~\gev$]{
        \includegraphics[width=0.45\textwidth]{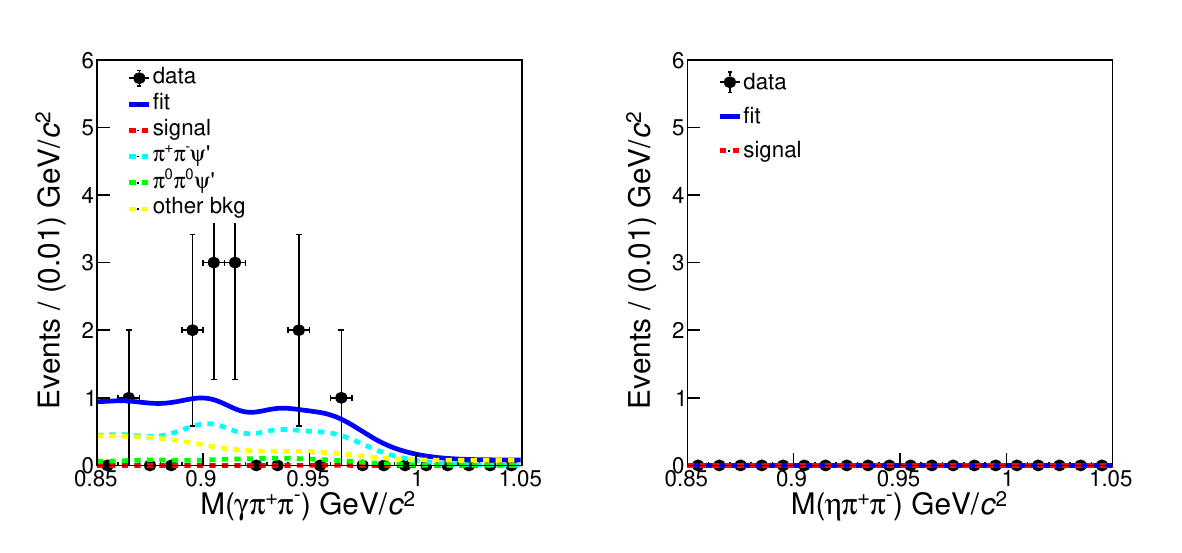}}
        \subfigure[~$4.68~\gev$]{
        \includegraphics[width=0.45\textwidth]{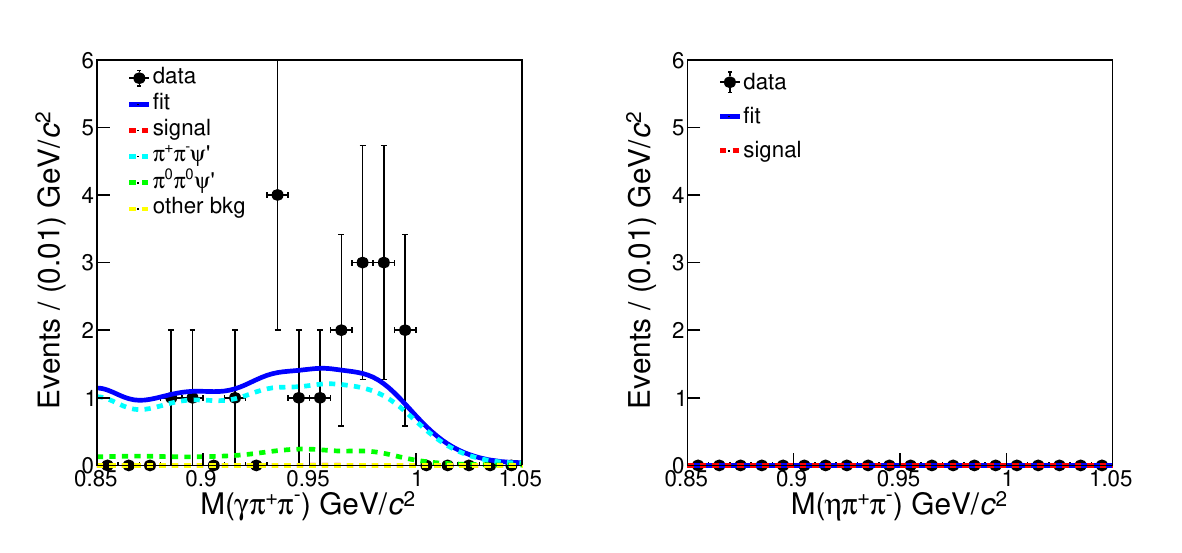}}
        \subfigure[~$4.70~\gev$]{
        \includegraphics[width=0.45\textwidth]{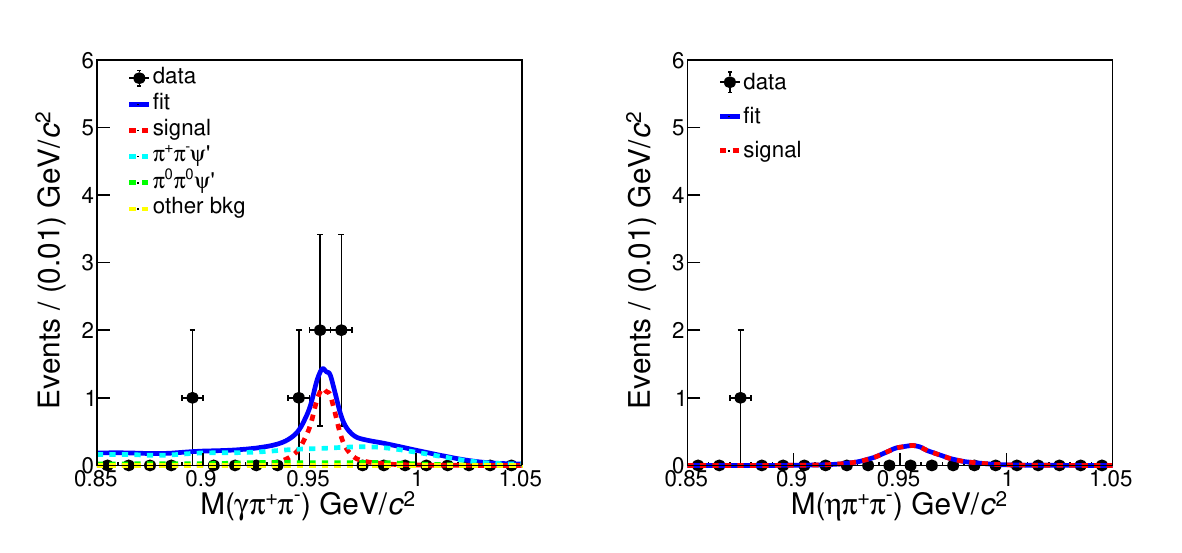}}
        \subfigure[~$4.74~\gev$]{
        \includegraphics[width=0.45\textwidth]{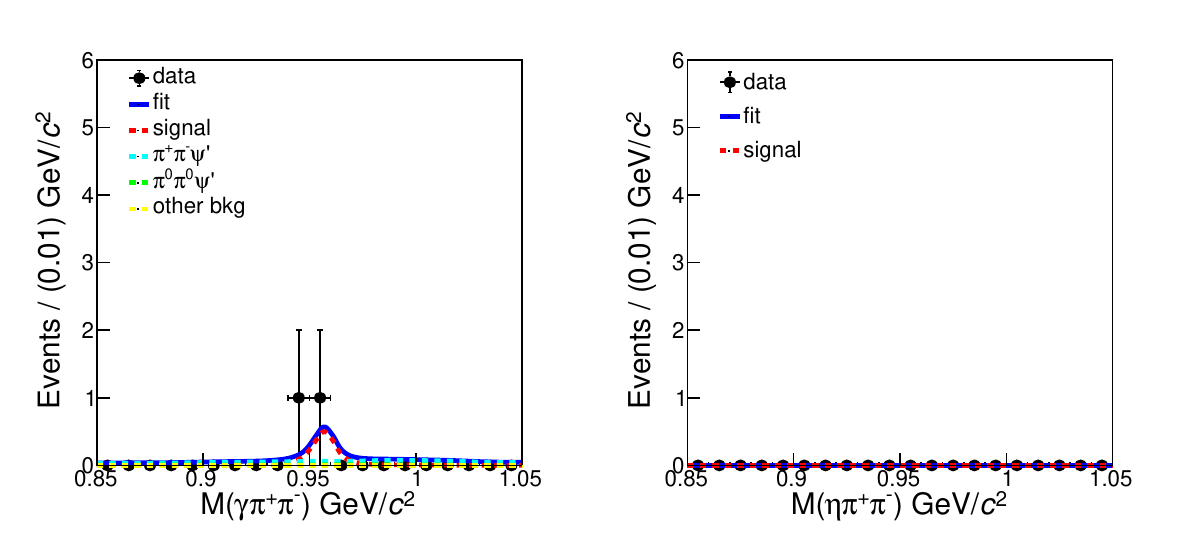}}
        \subfigure[~$4.75~\gev$]{
        \includegraphics[width=0.45\textwidth]{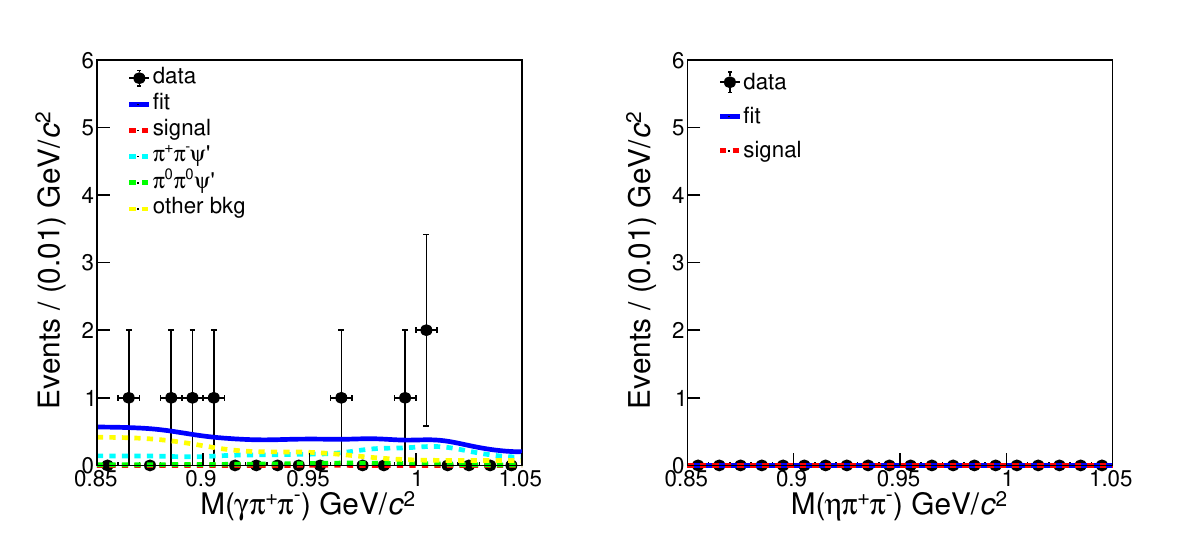}}
            \subfigure[~$4.78~\gev$]{
        \includegraphics[width=0.45\textwidth]{4780.pdf}}
            \subfigure[~$4.84~\gev$]{
        \includegraphics[width=0.45\textwidth]{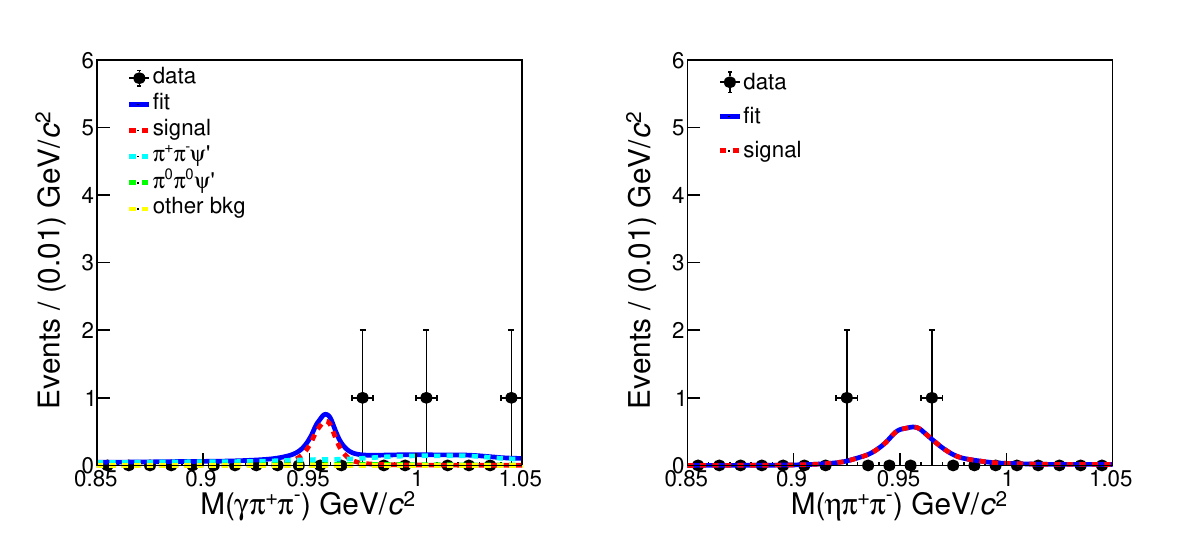}}
            \subfigure[~$4.914~\gev$]{

        \includegraphics[width=0.45\textwidth]{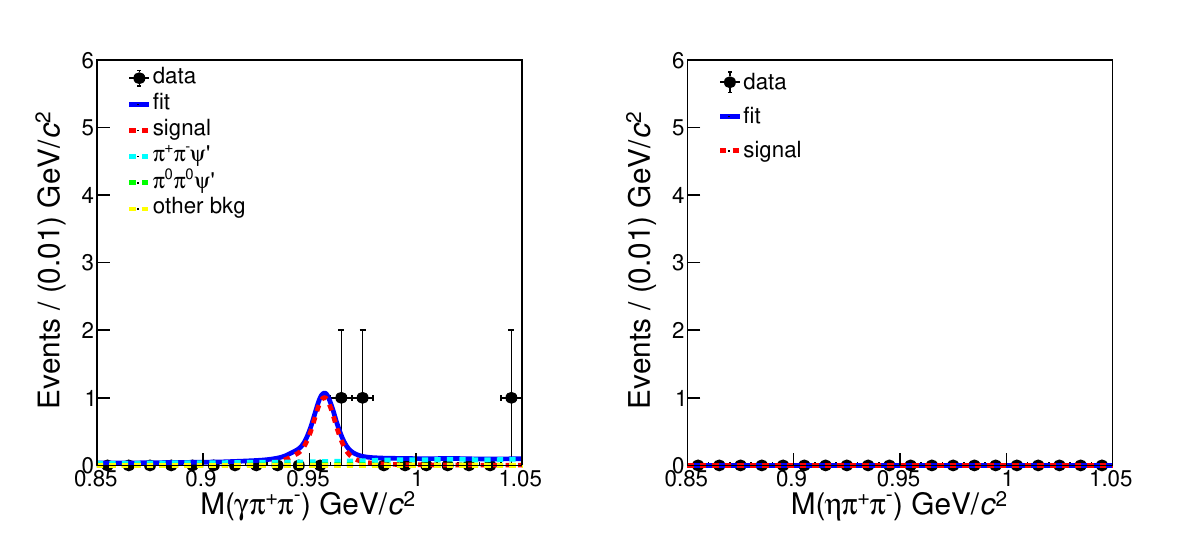}}
            \subfigure[~$4.946~\gev$]{
        \includegraphics[width=0.45\textwidth]{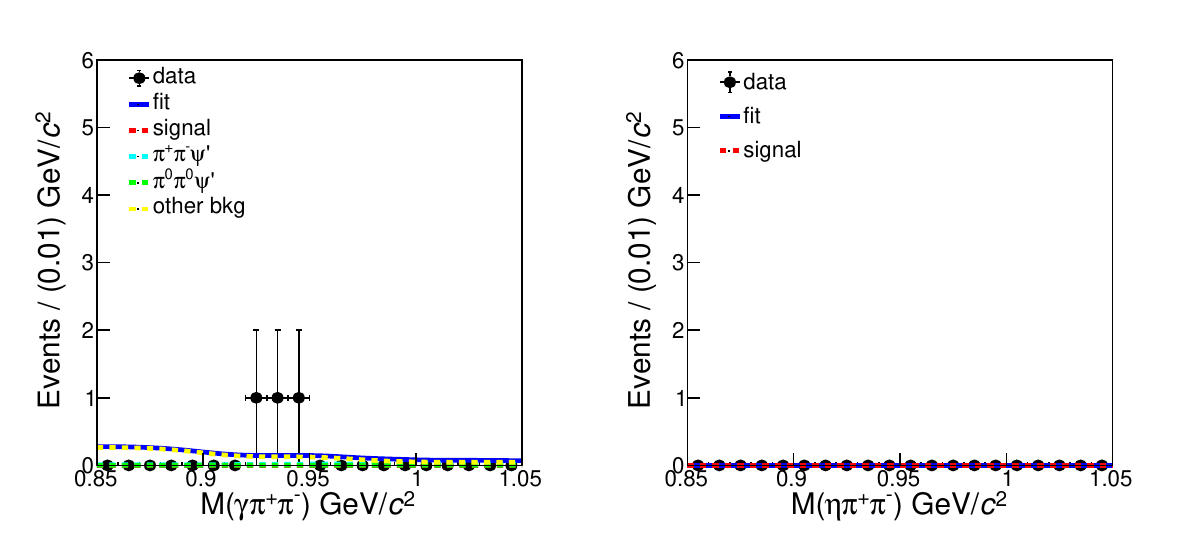}}
                    \subfigure[~Combined data]{
        \includegraphics[width=0.45\textwidth]{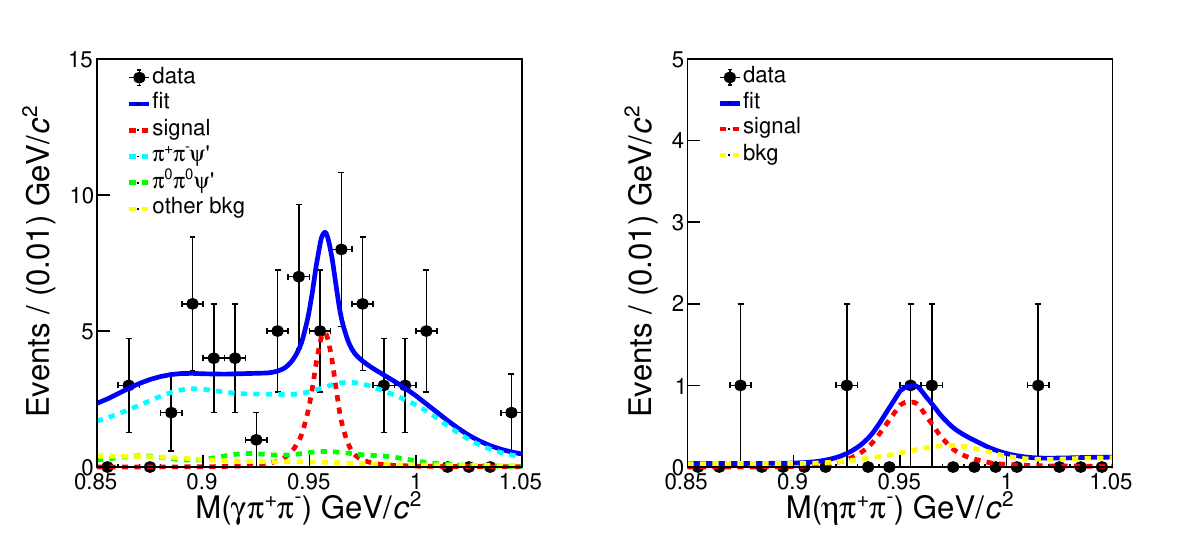}}
    \caption{
    Distributions of $M(\gamma\pp)$ for Mode I (left) and $M(\eta\pp)$ for Mode II (right) for data at all c.~m.~energies. 
    }
    \label{fig:fit2}
\end{figure*}


\end{document}